\documentclass[aps,twocolumn,prx,superscriptaddress,showpacs,tightenlines]{revtex4-1}
\usepackage{amsmath}
\usepackage{graphicx}
\usepackage{epsfig}
\usepackage{amsfonts}
\usepackage{color}
\usepackage{epstopdf}
\usepackage{hyperref}
\hypersetup{hypertex=true,
            colorlinks=true,
            linkcolor=blue,
            anchorcolor=blue,
            citecolor=blue}

\begin{document}

\newcommand{\nn}{\nonumber}
\newcommand{\ms}[1]{\mbox{\scriptsize #1}}
\newcommand{\msi}[1]{\mbox{\scriptsize\textit{#1}}}
\newcommand{\dg}{^\dagger}
\newcommand{\smallfrac}[2]{\mbox{$\frac{#1}{#2}$}}
\newcommand{\Tr}{\text{Tr}}
\newcommand{\ket}[1]{|#1\rangle}
\newcommand{\bra}[1]{\langle#1|}
\bibliographystyle{apsrev}
\newcommand{\pfpx}[2]{\frac{\partial #1}{\partial #2}}
\newcommand{\dfdx}[2]{\frac{d #1}{d #2}}
\newcommand{\half}{\smallfrac{1}{2}}
\newcommand{\s}{{\mathcal S}}
\newcommand{\jord}{\color{red}}
\newcommand{\kurt}{\color{blue}}

\title{ Dynamics of the driven open double two-level system and its entanglement generation}

\author{W. Ma }
\affiliation{School of Physics and Materials Engineering,
Dalian Nationalities University, Dalian 116600 China}

\author{X. L. Huang}
\affiliation{School of Physics and Electronic Technology,
Liaoning Normal University, Dalian 116029, China}

\author{S. L. Wu }
\email{slwu@dlnu.edu.cn}
\affiliation{School of Physics and Materials Engineering,
Dalian Nationalities University, Dalian 116600 China}

\date{\today}

\begin{abstract}

We investigate the dynamics of the driven open double two-level system by deriving a
driven-Markovian master equation based on the Lewis-Riesenfeld invariants theory.
The transitions induced by coupling to the heat reservoir occur between the instantaneous
eigenstates of the Lewis-Riesenfeld invariant. Therefore, different driving protocols
associated with corresponding Lewis-Riesenfeld invariants result in different open
system dynamics and symmetries. In particular, we show that, since the instantaneous
steady state of the driven double two-level system is one of eigenstates of the
Lewis-Riesenfeld invariant at ultra-low reservoir temperature, the inverse engineering
method based on the Lewis-Riesenfeld invariants has a good performance in
rapidly preparing quantum state of open quantum systems. As an example, a perfect
entangled state is generated by means of the inverse engineering method.

\end{abstract}

\maketitle

\section{Introduction}

The quantum control aims at manipulating the quantum system into desired
quantum states  or achieving certain quantum operations with satisfied fidelity
by using limited control manners. To improve the performance of the quantum
control, the dynamics of the quantum system has to be described as exactly as
possible. If the quantum system is isolated, its dynamics is governed by the
Schr\"{o}dinger equation, which results in a unitary evolution. Based on the
exact dynamical equation of the isolated systems, many effective methods
are proposed, such as the adiabatic control\cite{Kral2007,Hatomura2021},
the shortcuts to adiabaticity\cite{Odelin2019,Campo2019},
the optimal control\cite{Werschnik2007} and the Lyapunov
control\cite{Hou2012, Yi2009}. However, any system in
nature unavoidably couples to its surroundings. Therefore, the system
in the real world is never isolated, but an  open system\cite{Breuer2007}.
There are many different methods for describing dynamics of different
open quantum systems, such as the Markovian master equation\cite{Davies1974,Davies1978},
 the quantum state diffusion equation\cite{Flannigan2022}
 and the stochastic Schr\"{o}dinger equation\cite{Bouten2004}.

The master equation method gets special attention due to its concise expression
and clear physical meanings\cite{Breuer2007}.
To formulate the reduced dynamics of open quantum systems, one needs to
trace out the degree of freedom of the environment. In the original derivation
of the Markovian master equation, it is assumed that the system Hamiltonian is static.
The coupling with the environment induces transitions between the static eigenstates
of the system Hamiltonian\cite{Lindblad1976,Gorini1976}. But this master equation
cannot be used to describe the dynamics of the open systems with time-dependent
external drives. Many efforts have been devoted to develop the dynamical equation
of the driven open quantum systems. For the driving protocols fulfilling the adiabatic
condition, the master equation in the adiabatic limit is easy to be formulated
\cite{Davies1978,Albash2012,Childs2001,Kamleitner2013,Sarandy2004},
since the unitary transformation associated with the system Hamiltonian can be
given explicitly by the instantaneous eigenstates of the system Hamiltonian.
However, beyond the adiabatic limit, it is difficult to give a general master
equation without any restriction on the driving protocol\cite{Yamaguchi2017,
Dann2018,Potts2021}. Even for the simplest two-level system, deriving a non-adiabatic
Markovian master equation is a nontrivial task\cite{Dann2020}, let alone
multi-level systems.

Recently, based on the Lewis-Riesenfeld invariants (LRI) theory\cite{Lewis1968,Lewis1969},
a method of deriving the Markovian master equation for the driven open quantum
systems has been proposed, which is known as the driven Markovian master equation
(DMME)\cite{Wu2022}. The DMME can be used to describe the open system dynamics
with arbitrary control protocols under the Born-Markovian approximation.
In this paper, we derive the DMME for a driven double two-level system coupled to a
common heat reservoir. We show that the transition induced by the decoherence
occurs between the eigenstates of the LRI but not the system Hamiltonian's. And the
decoherence-free subspaces still emerge in the dynamics\cite{Karasik2008,Wu2017}.
Moreover, if the environment is a vacuum reservoir, the instantaneous steady state is
one of  eigenstates of the invariant. Therefore, the inverse engineering method based on the
Lewis-Riesenfeld invariants may present a high performance in this eigenstate\cite{Chen2010,Chen2011,Herrera2014}.
We verify this observation by proposing a protocol to generate a maximally entangled state with a perfect fidelity.
Since the decoherence always draw a quantum state into the instantaneous steady
state (the eigenstate used in the inverse engineering), our control protocol is also
robust to the imperfect initial state preparation. Hence, the inverse engineering method
is more robust to the decoherence than pervious predictions \cite{Jing2013}.

The paper is organized as follows. In Sec. \ref{sec:gf}, we briefly review the general formula
of the DMME for the driving protocol without the memory effect. Then, we derive the DMME for
the double two-level system with a time-dependent transverse field and scalar coupling in Sec.
\ref{sec:2qubits}, and present the corresponding adiabatic master equation and its instantaneous
steady state. In Sec. \ref{sec:REG}, we show that the inverse engineering scheme for the closed systems
works well if the driven double two-level system couples to a vacuum reservoir, and a maximally
entangled state can be generated with a perfect fidelity.  Finally, the conclusions are given in Sec.
\ref{sec:Conc}.

\section{The Driven Markovian Master Equation} \label{sec:gf}

We start by briefly reviewing the DMME based on the
LRI theory. Consider the dynamics of the composite system which is governed by
the Hamiltonian
\[H(t)=H_{\text{s}}(t)+H_{\text{B}}+H_{\text{I}}.\]
$H_{\text{s}}(t)$ stands for the system Hamiltonian. The reservoir is represented by the
Hamiltonian
\[
H_{\text{B}}=\sum_{k}\hbar\omega_{k}b_{k}^{\dagger}b_{k},
\]
in which $b_{k}$ and $\omega_k$ are the annihilation
operator and the eigen-frequency of the $k$-th mode of the reservoir.
The system-bath interaction Hamiltonian is given by
\[
H_{\text{I}}=\sum_{k}\textsl{g}_{k}A_{k}\otimes B_{k}.
\]
$A_{k}$ and $B_{k}$ are the Hermitian operators of the system and reservoir, respectively.
 $g_{k}$ stands for the coupling strength.

By assuming weak-system-bath coupling, the dynamics of the driven system is described
by the following Redfield master equation within the Born-Markovian approximation
\cite{Redfield1965},
\begin{eqnarray}
\partial_{t}\tilde{\rho}_{\text{s}}(t)\nonumber \\
=-\frac{1}{\hbar^{2}} & \int_{0}^{\infty}\text{d}s\,\text{Tr}_{\text{B}}
\left[\tilde{H}_{\text{I}}(t),\left[\tilde{H}_{\text{I}}(t-s),
\tilde{\rho}_{\text{s}}(t)\otimes\rho_{\text{B}}\right]\right],\label{eq:rfmequ}
\end{eqnarray}
where $\tilde{\rho}_{\text{s }}(t)$ is the reduced density matrix of the driven system in
the interaction representation, and a similar notation is applied for the other system
operators. For a system operator $A_k$, the corresponding operator in the interaction
picture can be connected by a unitary transformation as
\[
\tilde{A}_{k}(t)=\hat{\mathcal{U}}_{\text{s}}(t)A_{k}=
U_{\text{s}}^{\dagger}(t)A_{k}U_{\text{s}}(t).
\]
$U_{\text{s}}(t)$ describes the free dynamics of the system, which satisfies the Schr\"{o}dinger
equation with the system Hamiltonian
\begin{equation}
i\hbar\partial_{t}U_{\text{s}}(t)=H_{\text{s}}(t)U_{\text{s}}(t),\:U_{\text{s}}(0)=I.\label{eq:ueqn}
\end{equation}
This results in $U_{\text{s}}(t)=\mathcal{T}\exp\left(-i/\hbar\int_{0}^{t}\text{d}\tau\,
 H_{\text{s}}(\tau)\right)$ with the time-ordering operator $\mathcal{T}$.

The free dynamics of the system can be solved by means of the LRI theory.
The LRI   $I_{\text{s}}(t)$ for the systems with the Hamiltonian
$H_{\text{s}}(t)$  is a  Hermitian operator which obeys\cite{Lewis1969}
\begin{equation}
i\hbar\partial_{t}I_{\text{s}}(t)-\left[H_{\text{s}}(t),I_{\text{s}}(t)\right]=0,\label{eq:dIs}
\end{equation}
which is an equation in the  Schr\"{o}dinger picture. The quantum state of an
isolated system with the time-dependent Hamiltonian $H_{\text{s}}(t)$ can be
expressed in terms of the eigenstates of the LRI,
\begin{equation}
|\Psi(t)\rangle=\sum_{n=1}^{N}c_{n}\exp\left(i\alpha_{n}(t)\right)|\psi_{n}(t)\rangle.
\label{eq:psit}
\end{equation}
Here, $|\psi_{n}(t)\rangle$ is the n-th eigenstate of the LRI $I_{\text{s}}(t)$  with a real
constant eigenvalue $\lambda_{n}$, i.e., $I_{\text{s}}(t)|\psi_{n}(t)\rangle=\lambda_{n}
|\psi_{n}(t)\rangle$, $\{c_{n}\}$ are time-independent amplitudes, and the Lewis-Riesenfeld
phases are defined as \cite{Lewis1969}
\begin{equation}
\alpha_{n}(t)=\frac{1}{\hbar}\int_{0}^{t}\langle\psi_{n}(\tau)|\left(i\hbar\partial_{\tau}-
H_{\text{s}}(\tau)\right)|\psi_{n}(\tau)\rangle\,\text{d}\tau.\label{eq:lrp}
\end{equation}
Therefore, the solution of Eq.(\ref{eq:ueqn}) can be expressed by means of the eigenstates of
the LRI,
\begin{equation}
U_{\text{s}}(t)=\sum_{n}\exp\left(i\alpha_{n}(t)\right)|\psi_{n}(t)\rangle\langle\psi_{n}(0)|.
\label{eq:us}
\end{equation}
In the adiabatic limit, the changes of the eigenstates of the LRI can be neglected. Therefore,
the LRI and the system Hamiltonian must share common eigenstates, due to
$\left[H_{\text{s}}(t),I_{\text{s}}(t)\right]=0$.

By means of the explicit formula of the free evolution operator $U_{\text{s}}(t)$, the system
operator in the interaction picture reads
\begin{eqnarray}
\tilde{A}_{k}(t) & = & U_{\text{s}}^{\dagger}(t)A_{k}U_{\text{s}}(t)\nonumber \\
 & = & \sum_{n,m}\exp\left(i\theta_{mn}^{k}(t)\right)\xi_{mn}^{k}(t)\tilde F_{mn},\label{eq:Fj}
\end{eqnarray}
with
\begin{eqnarray}
\theta_{mn}^{k}(t)=\alpha_{n}(t)-\alpha_{m}(t)+\text{Arg}\left(\langle\psi_{m}(t)|
A_{k}|\psi_{n}(t) \rangle\right)
\label{eq:if}
\end{eqnarray}
and $\xi_{mn}^{k}(t)=|\langle\psi_{m}(t)|A_{k}|\psi_{n}(t)\rangle|$.
The time-independent operator $\tilde F_{mn}=|\psi_{m}(0)\rangle\langle\psi_{n}(0)|$
denotes one of the Lindblad operators in the interaction picture. The time-dependent coefficients satisfy
$\theta_{mn}^{k}(t),\:\xi_{mn}^{k}(t)\in\mathbb{R}$ and $\xi_{mn}^{k}(t)>0.$ Since
$\tilde{A}_{k}(t)$ are Hermitian operators, it yields
\begin{equation}
\tilde{A}_{k}(t)=\sum_{n',m'}\exp\left(-i\theta_{m'n'}^{k}(t)\right)\xi_{m'n'}^{k}(t)\tilde
F_{m'n'}^{\dagger}. \label{eq:Fjd}
\end{equation}
Any $F_{m'n'}^{\dagger}$ contains in the operator set $\left\{ F_{mn}\right\} $ which
expands the corresponding Hilbert-Schmidt space\cite{Petrosky1997}. Substituting
Eqs.(\ref{eq:Fj}) and (\ref{eq:Fjd}) into Eq.(\ref{eq:rfmequ}), the master equation reads
\begin{eqnarray*}
\partial_{t}\tilde{\rho}_{\text{s}} & (t)= & \frac{1}{\hbar^{2}}\sum_{k,k'}\sum_{m,m',n,n'}
\int_{0}^{\infty}\text{d}s\xi_{m'n'}^{k'}(t)\xi_{mn}^{k}(t-s)\textsl{g}_{k}\textsl{g}_{k'}\\
 & \times & \text{Tr}_{\text{B}}\left[\tilde{B}_{k'}(t)\tilde{B}_{k}(t-s)\rho_{B}\right]
 \text{e}^{i\left(\theta_{mn}^{k}(t-s)-\theta_{m'n'}^{k'}(t)\right)}\\
 & \times & \left(\tilde F_{mn}\tilde{\rho}_{\text{s}}(t)\tilde F_{m'n'}^{\dagger}-
 \tilde F_{m'n'}^{\dagger}\tilde  F_{mn}\tilde{\rho}_{\text{s}}(t)\right)+\text{H.c.},
\end{eqnarray*}
where H.c. denotes the Hermitian conjugated expression and $\tilde{B}_{k'}(t)$ is the
bath operator in the interaction picture.

For simplifying our discussion, we consider that the bath dynamics is fast compared to the driving rate
\cite{Dann2018}. In other words, the bath correlation decay time $\tau_{B}$ should
be much shorter than the non-adiabatic timescale $\tau_{d}$, which associates with
the change in the driving protocol. In such a case, the memory effect of the driving
can be neglected safely. For $s\in\left[0,\tau_{B}\right]$ and
$s\ll t$, $\xi_{mn}(t-s)$ and $\theta_{mn}(t-s)$ can be approximated by a polynomial
expansion in orders of $s$,
\begin{eqnarray*}
\xi_{mn}^k(t-s) & = & \xi_{mn}^k(t),\\
\theta_{mn}^k(t-s) & = & \theta_{mn}^k(t)-\partial_{t}\theta_{mn}^k(t)s\equiv\theta_{mn}^k(t)+
\alpha_{mn}^k(t)s,
\end{eqnarray*}
 which leads to the general formulism of the DMME  in the  interaction picture
\begin{eqnarray}
\partial_{t}\tilde{\rho}_{\text{s}} & (t)&=  \mathcal{\tilde{L}}(t)\tilde{\rho_{\text{s}}}\nonumber\\
&=&\sum_{k,k'}\sum_{m,m',n,n'}\xi_{m'n'}^{k'}(t)
\xi_{mn}^k(t)\text{e}^{i\left(\theta_{mn}^k(t)-\theta_{m'n'}^{k'}(t)\right)}\nonumber\\
 & \times & \Gamma_{kk'}(\alpha_{mn})\left(\tilde F_{mn}\tilde{\rho}_{\text{s}}(t)
 \tilde F_{m'n'}^{\dagger}-\tilde F_{m'n'}^{\dagger}\tilde F_{mn}\tilde{\rho}_{\text{s}}(t)\right)\nonumber\\
 & + & \text{H.c.},\label{eq:nsme}
\end{eqnarray}
with the one-side Fourier transforms of the correlation function of the bath operators
\begin{eqnarray}
\Gamma_{kk'}&(\alpha_{nm}) & = \frac{1}{\hbar^{2}}\textsl{g}_{k}\textsl{g}_{k'}
\label{eq:gammak}\\
 &  & \times\int_{0}^{\infty}\text{d}s\text{e}^{i\alpha_{mn}^ks}\text{Tr}_{\text{B}}
 \left[\tilde{B}_{k'}(t)\tilde{B}_{k}(t-s)\rho_{B}\right].\nonumber
\end{eqnarray}
$\alpha_{mn}^k(t)$ stands for the instantaneous frequency for the transition from
$\ket{\psi_n(t)}$ to $\ket{\psi_m(t)}$, and
$\mathcal {\tilde{L} }(t)$ denotes the Liouvillian superoperator in the interaction picture.

According to the non-secular master equation Eq.(\ref{eq:nsme}), the Lindblad operator
$\tilde F_{mn}$ denotes a transition from the state $\ket{\psi_{n}(0)}$ to another
one $\ket{\psi_{m}(0)}$. In other words, the transitions caused by the decoherence occur
between eigenstates of the LRI for the driven open quantum systems. Combining  Eqs.
(\ref{eq:lrp}) and (\ref{eq:if}), the instantaneous frequency $\alpha_{mn}^k$  can be
divided into three parts, i.e.,
\begin{eqnarray}
\alpha_{mn}^k&=&-\frac{1}{\hbar}\left(\langle\psi_{m}(t)|H_{\text{s}}(t)|\psi_{m}(t)\rangle-
\langle\psi_{n}(t)|H_{\text{s}}(t)|\psi_{n}(t)\rangle\right)\nonumber\\
&&+i\left(\langle\psi_{m}(t)|\partial_{t}|\psi_{m}(t)\rangle-
\langle\psi_{n}(t)|\partial_{t}|\psi_{n}(t)\rangle\right)\nonumber\\
&&-\partial_t \text{Arg}\left(\langle\psi_{m}(t)|A_{k}|\psi_{n}(t) \rangle\right).
\end{eqnarray}
The first term in the above equation attributes to a difference between the energy average values
of the eigenstates $\ket{\psi_{n}(t)}$ and $\ket{\psi_{m}(t)}$. The second term is a geometric
contribution from the time-dependent eigenstates, while the third term comes from the phase changing
rate in the transitions caused by the interaction Hamiltonian. In the adiabatic limit, the eigenstates
of the LRI are the eigenstates of the system Hamiltonian, and the adiabatic condition
must be satisfied. Thus, the last two terms have no contributions to the instantaneous frequency,
while the first term becomes the energy gap between the $n$-th and the $m$-th eigenstate of
the system Hamiltonian, which leads to the adiabatic master equation given in Ref. \cite{Albash2012}.

\section{The Driven Double Two-Level System} \label{sec:2qubits}

In this section, we present the DMME for a driven double two-level system which couples
with a common heat reservoir. Here, we consider that the driven double two-level system
Hamiltonian takes the form
\begin{equation}
H_{\text{s}}(t)=J\pi\sigma_{x}^{(1)}\sigma_{x}^{(2)}+f(t)\left(\sigma_{z}^{(1)}+\sigma_{z}^{(2)}\right),
\label{eq:Hs}
\end{equation}
where $J$ is the scalar coupling and $f(t)$ is a time-dependent modulation function. This is
the typical Hamiltonian of two coupled spins used in a scalar molecule in a NMR system
\cite{Oliveira2007,Maziero2013}.  The reservoir Hamiltonian reads
\[
H_{\text{B}}=\sum_{k}\hbar\omega_{k}b_{k}^{\dagger}b_{k},
\]
in which $b_{k}$ and $\omega_k$ are the annihilation
operator and the eigen-frequency of the $k$-th mode of the reservoir.
The interaction Hamiltonian only contains collective decay term, i.e.  $H_{\text{I}}=A
\otimes\sum_{k}\textsl{g}_{k}B$, where the system and bath operators are
\begin{equation}
A=\sigma_{x}^{(1)}+\sigma_{x}^{(2)},\:B_{k}=b_{k}^{\dagger}-b_{k}.\label{eq:bk}
\end{equation}
For the double two-level system with a Hamiltonian as in Eq.(\ref{eq:Hs}), the LRIs have been
explored before \cite{Herrera2014}, which read
\begin{eqnarray}
I_{\text{s}}(t) & = & g_{1}(t)\Sigma_{1}^{(1)}-g_{2}(t)\Sigma_{2}^{(1)}+g_{6}(t)\Sigma_{3}^{(1)}\nonumber \\
 & + & g_{3}(t)\Sigma_{1}^{(2)}+g_{4}(t)\Sigma_{2}^{(2)}-g_{5}(t)\Sigma_{3}^{(2)},\label{eq:Is}
\end{eqnarray}
with
\begin{eqnarray*}
\Sigma_{1}^{(1)} & = & \frac{\sigma_{z}^{(1)}+\sigma_{z}^{(2)}}{2},\,\Sigma_{2}^{(1)}=
-\frac{\sigma_{y}^{(1)}\sigma_{x}^{(2)}+\sigma_{x}^{(1)}\sigma_{y}^{(2)}}{2},\\
\Sigma_{3}^{(1)} & = & \frac{\sigma_{x}^{(1)}\sigma_{x}^{(2)}-\sigma_{y}^{(1)}\sigma_{y}^{(2)}}{2},
\end{eqnarray*}
and
\begin{eqnarray*}
\Sigma_{1}^{(2)} & = & \frac{\sigma_{x}^{(1)}\sigma_{x}^{(2)}+\sigma_{y}^{(1)}
\sigma_{y}^{(2)}}{2},\,\Sigma_{2}^{(2)}=\frac{\sigma_{z}^{(1)}-\sigma_{z}^{(2)}}{2},\\
\Sigma_{3}^{(2)} & =- & \frac{\sigma_{y}^{(1)}\sigma_{x}^{(2)}-\sigma_{x}^{(1)}
\sigma_{y}^{(2)}}{2}.
\end{eqnarray*}
The sets $\left\{ \Sigma_{i}^{(1)}\right\} _{i=1}^{3}$ and $\left\{ \Sigma_{i}^{(2)}
\right\} _{i=1}^{3}$ provide two independent \emph{su}(2) algebras with
 $\left[\Sigma_{i}^{(\alpha)},\Sigma_{j}^{(\alpha)}\right]=2i\varepsilon_{ijk}
\Sigma_{k}^{(\alpha)}$ for $\alpha=1,2$. $\varepsilon_{ijk}$ is the Levi-Civita
symbol. Inserting Eqs. (\ref{eq:Hs}) and (\ref{eq:Is}) into Eq.(\ref{eq:dIs}), we have
\begin{eqnarray}
\hbar\partial_{t}g_{1}(t) & = & 2\pi J(t)g_{2}(t),\nonumber \\
\hbar\partial_{t}g_{2}(t) & = & 4f(t)g_{6}(t)-2\pi J(t)g_{1}(t),\label{eq:geq}\\
\hbar\partial_{t}g_{3}(t) & = & 0,\nonumber \\
\hbar\partial_{t}g_{4}(t) & = & 2\pi J(t)g_{5}(t),\nonumber \\
\hbar\partial_{t}g_{5}(t) & = & -2\pi J(t)g_{4}(t),\nonumber \\
\hbar\partial_{t}g_{6}(t) & = & -4f(t)g_{2}(t).\nonumber
\end{eqnarray}
The eigenstates of the LRI (Eq.(\ref{eq:Is})) in the basis $\left\{ \ket{00},\ket{01},
\ket{10},\ket{11}\right\} $ are obtained after some simple algebra, which yields
\begin{eqnarray}
\ket{\psi_{1}(t)} & = & \left(0,-\cos\eta_{1}(t)\text{e}^{i\zeta_{1}(t)},\sin\eta_{1}(t),0\right)^{\text{T}},\label{eq:eigLRI}\\
\ket{\psi_{2}(t)} & = & \left(0,\sin\eta_{1}(t)\text{e}^{i\zeta_{1}(t)},\cos\eta_{1}(t),0\right)^{\text{T}},\nonumber \\
\ket{\psi_{3}(t)} & = & \left(-\cos\eta_{2}(t)\text{e}^{i\zeta_{2}(t)},0,0,\sin\eta_{2}(t)\right)^{\text{T}},\nonumber \\
\ket{\psi_{4}(t)} & = & \left(\sin\eta_{2}(t)\text{e}^{i\zeta_{2}(t)},0,0,\cos\eta_{2}(t)\right)^{\text{T}},\nonumber
\end{eqnarray}
with
\begin{eqnarray}
\sin\eta_{1}(t) & = & \sqrt{\frac{g_{3}^{2}(t)+g_{5}^{2}(t)}{2\,g_{4}(t)\,
\lambda_{1}+2\lambda_{1}^{2}}},\,\tan\zeta_{1}(t)=-\frac{g_{5}(t)}{g_{3}(t)},\nonumber\\
\sin\eta_{2}(t) & = & \sqrt{\frac{g_{6}^{2}(t)+g_{2}^{2}(t)}{2\,g_{1}(t)\,
\lambda_{3}+2\lambda_{3}^{2}}},\,\tan\zeta_{2}(t)=-\frac{g_{2}(t)}{g_{6}(t)}.\label{eq:zeta}
\end{eqnarray}
The corresponding eigenvalues are constants, which take the forms
\begin{eqnarray*}
\lambda_{1} & = & -\sqrt{g_{3}^{2}(t)+g_{4}^{2}(t)+g_{5}^{2}(t)},\\
\lambda_{2} & = & \sqrt{g_{3}^{2}(t)+g_{4}^{2}(t)+g_{5}^{2}(t)},\\
\lambda_{3} & = & -\sqrt{g_{1}^{2}(t)+g_{2}^{2}(t)+g_{6}^{2}(t)},\\
\lambda_{4} & = & \sqrt{g_{1}^{2}(t)+g_{2}^{2}(t)+g_{6}^{2}(t)}.
\end{eqnarray*}
Then, it is easy to obtain $A_{mn}=\bra{\psi_m(t)}A\ket{\psi_n(t)}$ by considering
Eqs.(\ref{eq:bk}) and (\ref{eq:eigLRI}),
\begin{eqnarray*}
A_{13} & = & \left(\sin\eta_{2}-\text{e}^{i\zeta_{2}}\cos\eta_{2}\right)
\left(\sin\eta_{1}-\text{e}^{-i\zeta_{1}}\cos\eta_{1}\right),\\
A_{14} & = & \left(\cos\eta_{2}+\text{e}^{i\zeta_{2}}\sin\eta_{2}\right)
\left(\sin\eta_{1}-\text{e}^{-i\zeta_{1}}\cos\eta_{1}\right),\\
A_{23} & = & \left(\sin\eta_{2}-\text{e}^{i\zeta_{2}}\cos\eta_{2}\right)
\left(\cos\eta_{1}+\text{e}^{-i\zeta_{1}}\sin\eta_{1}\right),\\
A_{24} & = & \left(\cos\eta_{2}+\text{e}^{i\zeta_{2}}\sin\eta_{2}\right)
\left(\cos\eta_{1}+\text{e}^{-i\zeta_{1}}\sin\eta_{1}\right),
\end{eqnarray*}
and the Lewis-Riesenfeld phases defined in Eq.(\ref{eq:lrp}) read
\begin{eqnarray*}
\alpha_{1} & = & -\frac{1}{\hbar}\int_{0}^{t}\text{d}\tau
\left(\hbar\partial_{\tau}\zeta_{1}\cos^{2}\eta_{1}+\pi J\sin2\eta_{1}\cos\zeta_{1}\right),\\
\alpha_{2} & = & -\frac{1}{\hbar}\int_{0}^{t}\text{d}\tau
\left(\hbar\partial_{\tau}\zeta_{1}\sin^{2}\eta_{1}-\pi J\sin2\eta_{1}\cos\zeta_{1}\right),\\
\alpha_{3} & = & -\frac{1}{\hbar}\int_{0}^{t}\text{d}\tau
\left(\hbar\partial_{\tau}\zeta_{2}\cos^{2}\eta_{2}-\pi J\sin2\eta_{2}\cos\zeta_{2}\right.\\
 &  & \left.+2f\cos2\eta_{2}\right),\\
\alpha_{4} & = & -\frac{1}{\hbar}\int_{0}^{t}\text{d}\tau
\left(\hbar\partial_{\tau}\zeta_{2}\sin^{2}\eta_{2}+\pi J\sin2\eta_{2}\cos\zeta_{2}\right.\\
 &  & \left.-2f\cos2\eta_{2}\right).
\end{eqnarray*}

Here, we consider that the double two-level system couples to a heat reservoir at temperature
$T_R$. The correlation functions of the reservoir satisfy
\begin{eqnarray*}
\text{Tr}_{\text{B}}\left[b_{k'}b_{k}^{\dagger}\rho_{B}\right] & = & \delta_{k'k}(1+N_{k}),\\
\text{Tr}_{\text{B}}\left[b_{k'}^{\dagger}b_{k}\rho_{B}\right] & = & \delta_{k'k}N_{k},\\
\text{Tr}_{\text{B}}\left[b_{k'}b_{k}\rho_{B}\right] & = & 0,\\
\text{Tr}_{\text{B}}\left[b_{k'}^{\dagger}b_{k}^{\dagger}\rho_{B}\right] & = &0,
\end{eqnarray*}
where $N_{k}=\left(\exp(\hbar\omega_{k}/\verb"k"T_R)-1\right)^{-1}$ denotes the Planck
distribution with the reservoir temperature $T_R$ and the Boltzmann's constant $\verb"k"$.
In continuum limit, the sum over $\textsl{g}_{k}^{2}$ can be replaced by an
integral
\[
\sum_{k}\textsl{g}_{k}^{2}\rightarrow\int_{0}^{\infty}\text{d}\omega_{k}J(\omega_{k})
\]
with the spectral density function $J(\omega_{k})$. Inserting Eq.(\ref{eq:bk}) into
Eq.(\ref{eq:gammak}), we obtain
\begin{eqnarray*}
\Gamma(\alpha_{mn}) &\equiv&\Gamma_{mn}^{(N)}\\
& = & \frac{1}{\hbar^{2}}\int_{0}^{\infty}\text{d}\omega_{k}J(\omega_{k})\left(N_{k}
\int_{0}^{\infty}\text{d}s\text{e}^{i\left(\omega_{k}+\alpha_{mn}\right)s}\right.\\
 &  & +(N_{k}+1)\int_{0}^{\infty}\text{d}s\text{e}^{-i\left(\alpha_{mn}-\omega_{k}\right)s}.
\end{eqnarray*}
Thus the Liouvillian superoperator $\mathcal{\tilde L}$  is
\begin{eqnarray}
\mathcal{\tilde L}\tilde{\rho}_{\text{s}}(t) & = & \sum_{m,m',n,n'}\xi_{m'n'}(t)
\xi_{mn}(t)\text{e}^{i\left(\theta_{mn}(t)-\theta_{m'n'}(t)\right)}\nonumber\\
 & \times & \Gamma_{mn}^{(N)}\left(\tilde F_{mn}\tilde{\rho}_{\text{s}}(t)
\tilde  F_{m'n'}^{\dagger}-\tilde F_{m'n'}^{\dagger}\tilde F_{mn}\tilde{\rho}_{\text{s}}(t)\right)\nonumber\\
 & + & \text{H.c.}.\label{eq:brequ}
\end{eqnarray}
{In order to guarantee the complete positivity of the driven Markovian master equation,} we
neglect fast oscillating terms in above equation, which satisfy $\theta_{mn}(t)=\theta_{m'n'}(t)$.
Under the secular approximation, we have
\begin{eqnarray*}
\mathcal{\tilde L}\tilde{\rho}_{\text{s}}(t) & = & \sum_{m,m',n,n'}^{\theta_{mn}=\theta_{m'n'}}\xi_{m'n'}(t)
\xi_{mn}(t)\Gamma_{mn}^{(N)}\\
 & \times & \left(\tilde F_{mn}\tilde{\rho}_{\text{s}}(t)\tilde F_{m'n'}^{\dagger}
 -\tilde F_{m'n'}^{\dagger}\tilde F_{mn}\tilde{\rho}_{\text{s}}(t)\right)+\text{H.c.}.
\end{eqnarray*}
If $\theta_{mn}(t)=\theta_{m'n'}(t)$ only for $m=m'$ and $n=n'$, it yields
\begin{eqnarray*}
\mathcal{\tilde L}\tilde{\rho}_{\text{s}}= \sum_{mn}\xi_{mn}^{2}\Gamma_{mn}^{(N)}
\left(2\tilde F_{mn}\tilde{\rho}_{\text{s}}
\tilde F_{mn}^{\dagger}-\left\{ \tilde F_{mn}^{\dagger}\tilde F_{mn},
\tilde{\rho}_{\text{s}}\right\} \right).
\end{eqnarray*}
{Here, we need to state that, since the instantaneous frequencies
$\alpha_{mn}(t)$ are time-dependent, the secular approximation may
not be satisfied. But if the Lindblad superoperator given by
Eq.(\ref{eq:brequ}) presents some special symmetries, the partial secular
approximation can be used to reduce the complexity of the master
equation while ensuring the  complete positivity of the master equation
\cite{Giovannetti2019,Cattaneo2020}.}

For $\Gamma^{(N)}(\alpha_{mn})$, by making use of the formula
\[
\int_{0}^{\infty}\text{d}s\text{e}^{-i\varepsilon s}=\pi\delta(\varepsilon)
-i\text{P}\frac{1}{\varepsilon}
\]
with the Cauchy principal value P, we finally arrive at
\[
\Gamma_{mn}^{(N)}=\frac{1}{2}\gamma(\alpha_{mn})+iS(\alpha_{mn}),
\]
where we introduce the quantities
\[
\gamma(\alpha_{mn})=\gamma_{0}(\alpha_{mn})\left(N(\alpha_{mn})+1\right)
\]
and
\[
S(\alpha_{mn})=\text{P}\left[\int_{0}^{\infty}\text{d}\omega_{k}
\frac{J(\omega_k)}{\hbar^2}\left[\frac{N(\omega_{k})+1}
{\alpha_{mn}-\omega_{k}}+\frac{N(\omega_{k})}{\alpha_{mn}+\omega_{k}}\right]\right]
\]
with $\gamma_{0}(\alpha_{mn})=2\pi\hbar^{-2}J(\alpha_{mn}).$  Since the
Planck distribution satisfies $N(-\alpha_{mn})=-\left(N(\alpha_{mn})+1\right),$ the
master equation in the interaction picture  can be written as
\begin{eqnarray}
\mathcal{\tilde L}\tilde{\rho}_{\text{s}}(t)=-\frac{i}{\hbar}\left[\tilde H_{\text{LS}}(t),
\tilde{\rho}_{\text{s}}(t)\right]+\mathcal{D}^{(N)}\tilde{\rho}_{\text{s}}(t),\label{eq:imeq}
\end{eqnarray}
 where
 \begin{eqnarray}
 \tilde H_{\text{LS}}=\sum_{mn}\hbar S(\alpha_{mn})\xi_{mn}^{2}\tilde F_{mn}^{\dagger}
\tilde F_{mn} \label{eq:lmh}
\end{eqnarray}
 is the Lamb shift and the Stark shift. These shifts are induced by the fluctuations of the common
heat reservoir and the dissipator takes the form
\begin{eqnarray*}
\mathcal{D}^{(N)}\tilde{\rho}_{\text{s}} & = &\sum_{\alpha_{mn}>0}\xi_{mn}^{2}\gamma_{0}(\alpha_{mn})
 \left[\left(N(\alpha_{mn})+1\right)\right. \\&  &\times\left.\left(\tilde F_{mn}\tilde{\rho}_{\text{s}}\tilde F_{mn}^{\dagger}
 -\frac{1}{2}\left\{ \tilde F_{mn}^{\dagger}\tilde F_{mn},\tilde{\rho}_{\text{s}}\right\} \right)\right.\\
 &  &\left. +N(\alpha_{mn})\left(\tilde F_{mn}^{\dagger}\tilde{\rho}_{\text{s}}\tilde F_{mn}-\frac{1}{2}
 \left\{ \tilde F_{mn}\tilde F_{nm}^{\dagger},\tilde{\rho}_{\text{s}}\right\} \right)\right].
\end{eqnarray*}
Transforming back to the Schr\"{o}dinger picture, we finally arrive at the DMME,
\begin{eqnarray}
\partial_{t}\rho_{\text{s}} & = & \mathcal{L}(t)\rho_{\text{s}}\nonumber\\
 & = & -\frac{i}{\hbar}\left[H_{\text{s}}(t)+H_{\text{LS}}(t),\rho_{\text{s}}(t)\right]+
 \sum_{\alpha_{mn}>0}\xi_{mn}^{2}\gamma_{0}(\alpha_{mn})\nonumber\\
& \times& \left(\left[N(\alpha_{mn})+1\right]\left(F_{mn}\tilde{\rho}_{\text{s}}F_{mn}^{\dagger}
-\frac{1}{2}\left\{ F_{mn}^{\dagger}F_{mn},\tilde{\rho}_{\text{s}}\right\} \right)\right.\nonumber\\
 & + &\left. N(\alpha_{mn})\left(F_{mn}^{\dagger}\tilde{\rho}_{\text{s}}F_{mn}-\frac{1}{2}
 \left\{ F_{mn}F_{mn}^{\dagger},\tilde{\rho}_{\text{s}}\right\} \right)\right),\label{eq:smeq}
\end{eqnarray}
with the time-dependent Lindblad operators $F_{mn}=U_{\text{s}}(t)\tilde F_{mn}U_{\text{s}}^{\dagger}(t)$
and the Lamb shift $ H_{\text{LS}}=\sum_{mn}\hbar S(\alpha_{mn})\xi_{mn}^{2} F_{mn}^{\dagger}
 F_{mn}$ .

\subsection{The Adiabatic Limit}

In the adiabatic limit, the corresponding LRIs satisfy $\left[H_{\text{s}}(t),
I_{\text{s}}(t)\right]=0$, and share the same eigenstates to the system Hamiltonian.
According to Eq.(\ref{eq:geq}), if $\partial_{t}g_{i}(t)=0$, it yields $g_{6}=\pi Jg_{1}/2f$,
$g_{3}(t)=g_{3}(0)$ and $g_{i}=0$ for $i\neq1,3,6$. Thus, we obtain the eigenstates
of the system Hamiltonian Eq.(\ref{eq:Hs}) from Eq.(\ref{eq:eigLRI}) immediately.
We can verify the following eigen-equation
\[
H_{\text{s}}(t)\ket{\psi_{n}(t)}=\epsilon_{n}(t)\ket{\psi_{n}(t)},
\]
with the eigenvalues of the system Hamiltonian $\epsilon_{1,2}(t)=\mp\pi J,\:
\epsilon_{3,4}(t)=\mp\sqrt{(\pi J)^{2}+4f^{2}}$. In such a case, the propagator can be
represented in terms of the instantaneous eigenstates of the system Hamiltonian as
in Eq.(\ref{eq:us}). The phases in the propagator become  a sum of the geometric
phases and the dynamical phases. According to Eq.(\ref{eq:eigLRI}), we write down the
eigenstates of the system Hamiltonian  in the adiabatic limit with
$\zeta_{1}=0$, $\eta_{1}=\pi/4$, and
\begin{eqnarray}
\zeta_{2}&=&0,\nonumber\\
\eta_{2}&=&\arccos\left(\frac{\sqrt{2}}{2}\sqrt{\frac{\sqrt{\pi^{2}\,J^{2}+4\,f^{2}}-2\,f}
{\sqrt{\pi^{2}\,J^{2}+4\,f^{2}}}}\right).\label{eq:eta2}
\end{eqnarray}
Thus, the nonzero expansion coefficients in Eq.(\ref{eq:Fj}) are
\begin{eqnarray}
\xi_{23} & = & \xi_{32}=|\sqrt{2}\left(\cos\eta_{2}-\sin\eta_{2}\right)|\nonumber\\
 & = &\left|\frac{\left(\pi J+2f-\sqrt{\pi^{2}J^{2}+4f^{2}}\right)}
 {\sqrt{\sqrt{\pi^{2}\,J^{2}+4\,f^{2}}\left(\sqrt{\pi^{2}\,J^{2}+4\,f^{2}}-2f\right)}}\right|,\label{eq:xi23}\\
\xi_{24} & = & \xi_{42}=|\sqrt{2}\left(\cos\eta_{2}+\sin\eta_{2}\right)|\nonumber\\
 & =&\left| \frac{\left(\pi J+2f+\sqrt{\pi^{2}J^{2}+4f^{2}}\right)}
 {\sqrt{\sqrt{\pi^{2}J^{2}+4\,f^{2}}\left(\sqrt{\pi^{2}\,J^{2}+4\,f^{2}}+2f\right)}}\right|.\nonumber
\end{eqnarray}
Due to $\zeta_{1}=\zeta_{2}=0$, the geometric phase vanishes in $\alpha_{mn}$, so that
the phase in Eq.(\ref{eq:Fj}) reads
\begin{eqnarray*}
\theta_{23} & = & \alpha_{3}-\alpha_{2}\\
 & = & \frac{1}{\hbar}\int_{0}^{t}\text{d}\tau\left(
 \sqrt{\left(\pi J(\tau)\right)^{2}+4f^{2}(\tau)}+\pi J(\tau)\right),\\
\theta_{24} & = & \alpha_{4}-\alpha_{2}\\
 & = & -\frac{1}{\hbar}\int_{0}^{t}\text{d}\tau\left(
 \sqrt{\left(\pi J(\tau)\right)^{2}+4f^{2}(\tau)}-\pi J(\tau)\right),
\end{eqnarray*}
and $\theta_{mn}=-\theta_{nm}$, which leads to the instantaneous frequency vias $\alpha_{mn}=
-\partial_t \theta_{mn}(t)$,
\begin{eqnarray*}
\alpha_{23} & =- & \frac{1}{\hbar}\left(\sqrt{\left(\pi J(t)\right)^{2}+4f^{2}(t)}+\pi J(t)\right),\\
\alpha_{24} & = & \frac{1}{\hbar}\left(\sqrt{\left(\pi J(t)\right)^{2}+4f^{2}(t)}-\pi J(t)\right),
\end{eqnarray*}
and $\alpha_{mn}=-\alpha_{mn}$, respectively. No matter $J(t)$ is positive or negative, $\alpha_{32}$
and $\alpha_{24}$ are always positive. There are two Lindblad operators involved in Eq.(\ref{eq:smeq}),
 i.e., $F_{32}(t)=\exp\left(i\theta_{32}(t)\right)\ket{\psi_{3}(t)}\bra{\psi_{2}(t)}$
and $F_{24}(t)=\exp\left(i\theta_{24}(t)\right)\ket{\psi_{2}(t)}\bra{\psi_{4}(t)}$.

\subsection{The Instantaneous Steady State}

The instantaneous steady state $\tilde{\rho}_{\text{ss}}$ of the driven double two-level system in
the interaction picture satisfies $\tilde{\mathcal{L}}\tilde{\rho}_{\text{ss}}=0$\cite{Kraus2008}, which
can be expanded by the eigenstates of the dynamical invariants Eq.(\ref{eq:eigLRI})
at $t=0$,
\[
\tilde{\rho}_{\text{ss}}=\sum_{i,j}\rho_{ij}\ket{\psi_{i}(0)}\bra{\psi_{j}(0)}.
\]
{By  substituting the instantaneous steady state to Eq.(\ref{eq:imeq}) and considering the steady state
condition $\mathcal{\tilde L}\tilde{\rho}_{\text{ss}}=0$, it yields
\begin{eqnarray*}
&&-i\sum_{i,j}\sum_{m}\left(S(\alpha_{mi})\xi_{mi}^{2}-S(\alpha_{mj})\xi_{mj}^{2}\right)\rho_{ij}|\psi_{i}(0)\rangle\langle\psi_{j}(0)|\nonumber\\
&&+ \sum_{\alpha_{mn}>0}\xi_{mn}^{2}\gamma_{0}(\alpha_{mn})\left[\left(N_{mn}+1\right)\left(\rho_{mm}
 \ket{\psi_{n}(0)}\bra{\psi_{n}(0)}\right.\right.\\
 &  &\left. -\frac{1}{2}\sum_{i}\left(\rho_{mi}\ket{\psi_{m}(0)}\bra{\psi_{i}(0)}
 -\rho_{im}\ket{\psi_{i}(0)}\bra{\psi_{m}(0)}\right)\right)\\
 &  & +N_{mn}\left(\rho_{nn}\ket{\psi_{m}(0)}\bra{\psi_{m}(0)}\right.\\
 &  &\left. \left.-\frac{1}{2}\sum_{i}\left(\rho_{ni}(t)\ket{\psi_{n}(0)}\bra{\psi_{i}(0)}
 -\rho_{in}\ket{\psi_{i}(0)}\bra{\psi_{n}(0)}\right)\right)\right]\\
 &&=0.
\end{eqnarray*}
Here, we denote $N_{mn}\equiv N(\alpha_{mn})$ . }In the adiabatic limit, the Lindblad
operators $F_{32}$ and $F_{24}$ are survived, so that we have
\begin{eqnarray*}
0 & = & \xi_{32}^{2}\gamma_{0}(\alpha_{32})\left(\left(N_{32}+1\right)
\rho_{22}-N_{32}\rho_{33}\right)\\
 &  & +\xi_{24}^{2}\gamma_{0}(\alpha_{24})\left(-\left(N_{24}+1\right)
 \rho_{44}+N_{24}\rho_{22}\right),\\
0 & = & N_{32}\rho_{33}-\left(N_{32}+1\right)\rho_{22},\\
0 & = & \left(N_{24}+1\right)\rho_{44}-N_{24}\rho_{22}.
\end{eqnarray*}
As shown, a subspace with the basis $\ket{\psi_{1}(t)}$ decouples to the other parts
of the Hilbert space. Thus, if $\ket{\psi_{1}(t)}$ is not populated, the steady state has to
satisfy the following normalized condition
\[
\rho_{22}+\rho_{33}+\rho_{44}=1.
\]
 Immediately, we obtain the diagonal elements of the instantaneous steady state under the
 adiabatic limits,
\begin{eqnarray}
\rho_{22} & = & \frac{N_{32}(N_{24}+1)}{2N_{24}+N_{32}+3N_{24}N_{32}+1},\nonumber\\
\rho_{33} & = & \frac{(N_{24}+1)(N_{32}+1)}{2N_{24}+N_{32}+3N_{24}N_{32}+1},\nonumber\\
\rho_{44} & = & \frac{N_{24}N_{32}}{2N_{24}+N_{32}+3N_{24}N_{32}+1}.\label{eq:steadys}
\end{eqnarray}
And all of off-diagonal elements are trivial, i.e.,
\[\rho_{23}  = \rho_{32}=\rho_{24}=\rho_{42}=0.\]

{If the reservoir is vacuum ($N_{mn}=0$ for all $mn$), we obtain $\rho_{22}=\rho_{44}=0$
and $\rho_{33}=1$.  In other words, the instantaneous state is a pure state in the interaction
picture, i.e., $\tilde\rho_{\text{ss}}=\ket{\psi_{3}(0}\bra{\psi_{3}(0)}$. Thus the steady state
in the Schr\"{o}dinger must be a time-dependent pure state $\rho_{\text{ss}}(t)=\ket{\psi_{3}(t}
\bra{\psi_{3}(t)}$, due to $\rho_{\text{ss}}(t)=U_{\text{s}}(t)\tilde\rho_{\text{ss}}U_{\text{s}}^\dagger(t)$.
 In fact, $\ket{\psi_{3}(0)}$ is the dark state for the DMME with zero reservoir temperature
in the interaction picture. In order to show this, we consider the criteria of
the dark state of open quantum systems given in Ref.\cite{Kraus2008}. The theorem states that,
 for a Liouvillian superoperator $\tilde{\mathcal {L}}$ defined as in Eq.(\ref{eq:imeq}),  $\tilde{\mathcal {L}}\ket{\phi}
\bra{\phi}=0$ will be satisfied, if and only if the following two conditions are fulfilled: (i) $(-i \tilde H_{\text{LS}}+
\sum_{mn}\xi_{mn}^{2}\gamma_{0}\tilde F_{mn}^{\dagger}\tilde F_{mn})\ket{\phi}=\lambda\ket{\phi}$
for some $\lambda\in \mathbb{C}$; (ii)$\tilde F_{mn}\ket{\phi}=\lambda_{mn} \ket{\phi}$ for
some $\lambda_{mn}\in\mathbb{C}$ with $\sum_{mn}\xi_{mn}^{2}\gamma_{0}|\lambda_{mn}|^2
=\text{Re}(\lambda)$, where $\text{Re}(x)$ denotes the real part of $x$. For the driven double two-level system,
there are two Lindblad operators involved in the adiabatic master equation at zero reservoir temperature,
 i.e., $\tilde F_{32}=\ket{\psi_3(0)}\bra{\psi_2(0)}$ and $\tilde F_{24}=\ket{\psi_2(0)}\bra{\psi_4(0)}$,
 which yields $\tilde F_{32}\ket{\psi_{3}(0)}=\tilde F_{24}\ket{\psi_{3}(0)}=0$.
According to Eq. (\ref{eq:lmh}),  the Lamb shift Hamiltonian reads
\begin{eqnarray*}
 \tilde H_{\text{LS}}&&=\hbar S(\alpha_{32})\xi_{32}^{2}\ket{\psi_2(0)}\bra{\psi_2(0)}
\nonumber\\&&+\hbar S(\alpha_{24})\xi_{24}^{2}\ket{\psi_4(0)}\bra{\psi_4(0)},
\end{eqnarray*}
which results in $ \tilde H_{\text{LS}}\ket{\psi_{3}(0)}=0$. Therefore, it is obvious that $\ket{\psi_{3}(0)}$
is the dark state of the adiabatic master equation at zero reservoir temperature.}

Since the eigenstate $\eta_{2}$ is time-dependent,
we can use this pure instantaneous steady state to generate an entangle state
$(\ket{00}-\ket{11})/\sqrt{2}$ by means of the adiabatic engineering protocol
\cite{Wu2017,Sarandy2005,Venuti2016}. On the other hand, the eigenstate $\ket{\psi_{1}}$
of the LRI decouples to the other part of the Hilbert space. Therefore, there is a one-dimensional
decoherence-free subspace in this model\cite{Karasik2008,Wu2017,Altepeter2004}. The two-dimensional
decoherence-free subspace appears only if there is no scalar coupling (or the transverse field),
i.e., $J=0$ (or $f=0$) . At this time, it yields $\xi_{32}=0$ (see Eq.(\ref{eq:xi23})) and
$\eta_2=\pi/2$ ($\eta_2=\pi/4$),
so that $\ket{\psi_3(t)}$ decouples to the other parts of the Hilbert space. Since $\eta_2$ is
time-independent, $\ket{00}$ ($(\ket{00}-\ket{11})/\sqrt{2}$) will be another dimension of the
decoherence free subspace\cite{Qin2015}. This discussion is also held true for the finite
temperature reservoir.

\section{Rapid Entanglement state Generation} \label{sec:REG}

In this section, we show that the inverse engineering method works well when the driven
double two-level system couples to a common vacuum reservoir, i.e., $N_{mn}=0$ in Eq.(\ref{eq:imeq}).
Here, we generate an entanglement state $(\ket{00}-\ket{11})/\sqrt{2}$ by means of the
instantaneous steady state of the DMME, which belongs to a 2-dimensional decoherence-free
subspace at $f=0$.  Firstly, it can be observed from Eq.(\ref{eq:geq}) that $g_{3},\:g_{4}$
and $g_{5}$ decouple to the others. Since the scalar coupling satisfies $J\neq0$ in most of the time,
we consider a time-independent ansatz for $g_i$ for $i=3,4,5$, i.e.,  $g_{4}(t)=g_{5}(t)=0$ and
$g_{3}(t)=g_{3}(0)\neq 0$, which corresponds to $\eta_{1}=\pi/4$ and $\zeta_{1}=0$. Hence,
we have $A_{13}=A_{14}=A_{12}=A_{34}=0$, and
\begin{eqnarray*}
A_{23} & = & \sqrt{2\left(1-\sin2\eta_{2}\cos\zeta_{2}\right)}\text{e}^{i\varphi_{23}},\\
A_{24} & = & \sqrt{2\left(1+\sin2\eta_{2}\cos\zeta_{2}\right)}\text{e}^{i\varphi_{24}},
\end{eqnarray*}
with
\begin{eqnarray*}
\tan\varphi_{23} & = & -\frac{\cos\eta_{2}\sin\zeta_{2}}{\sin\eta_{2}-\cos\eta_{2}\cos\zeta_{2}},\\
\tan\varphi_{24} & = & \frac{\sin\eta_{2}\sin\zeta_{2}}{\cos\eta_{2}+\sin\eta_{2}\cos\zeta_{2}}.
\end{eqnarray*}
Via this parameters' setting, $\ket{\psi_{1}(t)}$ decouples to the other part of the Hilbert space.
It is straight forward to show
\begin{eqnarray*}
\xi_{23} & = & \sqrt{2\left(1-\sin2\eta_{2}\cos\zeta_{2}\right)},\\
\xi_{24} & = & \sqrt{2\left(1+\sin2\eta_{2}\cos\zeta_{2}\right)},
\end{eqnarray*}
 and
\begin{eqnarray*}
\theta_{23} & = & \alpha_{3}-\alpha_{2}+{\varphi}_{23},\\
\theta_{24} & = & \alpha_{4}-\alpha_{2}+{\varphi}_{24},
\end{eqnarray*}
 which results in
\begin{eqnarray*}
\alpha_{23} & = & \frac{1}{\hbar}\left(\hbar\dot{\zeta}_{2}\cos^{2}\eta_{2}+2f\cos2\eta_{2}\right.\\
 &  & \left.-\pi J\left(\sin2\eta_{2}\cos\zeta_{2}+1\right)\right)\\
 &  & -\frac{\dot{\zeta}_{2}\left(1+\cos2\eta_{2}-\sin2\eta_{2}\cos\zeta_{2}\right)+2
 \mathrm{\dot{\eta}_{2}}\sin\zeta_{2}}{2\,\sin2\eta_{2}\cos\zeta_{2}-2},\\
\alpha_{24} & = & \frac{1}{\hbar}\left(\hbar\dot{\zeta}_{2}\sin^{2}\eta_{2}-2f\cos2\eta_{2}\right.\\
 &  &+\left.\pi J\left(\sin2\eta_{2}\cos\zeta_{2}-1\right)\right)\\
 &  & +\frac{\dot{\zeta}_{2}\left(1-\cos2\eta_{2}+\sin2\eta_{2}\cos\zeta_{2}\right)+
 2\dot{\eta}_{2}\sin\zeta_{2}}{2\,\sin2\eta_{2}\cos\zeta_{2}+2}.
\end{eqnarray*}
Here, the dot denotes the derivative with respect to $t$.
 If $\eta_{2}$ and $\zeta_{2}$ do not change too radically, positive $\alpha_{32}$ and
 $\alpha_{24}$ can be ensured. Thus, the DMME in the Schr\"{o}dinger picture for the
double two-level system coupling to a common vacuum reservoir can be written as
\begin{eqnarray}
\partial_{t}{\rho}_{\text{s}}&=&-\frac{i}{\hbar}\left[H_{\text{s}}(t),\rho_{\text{s}}\right]\nonumber\\
 & +&\gamma_{32}\left(F_{32}(t){\rho}_{\text{s}}F_{32}^{\dagger}(t)-\frac{1}{2}
 \left\{ F_{32}^{\dagger}(t)F_{32}(t),{\rho}_{\text{s}}\right\} \right)\nonumber\\
 & + &\gamma_{24}\left(F_{24}(t){\rho}_{\text{s}}F_{24}^{\dagger}(t)-\frac{1}{2}
 \left\{ F_{24}^{\dagger}(t)F_{24}(t),{\rho}_{\text{s}}\right\} \right),\nonumber\\
 \label{eq:meq2}
\end{eqnarray}
where $\gamma_{32}= \xi_{32}^{2}\gamma_{0}(\alpha_{32})$ and
$\gamma_{24}= \xi_{24}^{2}\gamma_{0}(\alpha_{24})$
 are decoherence rates correspondingly. {Here the Lamb shift Hamiltonian is discarded, because
  $H_{\text{LS}}(t)$ does not affect the instantaneous steady state engineering
 evidently. The detailed discussion can be found in Append \ref{A1}.}

As we see, the DMME is similar to the adiabatic one, except that $\{\ket{\psi_{i}}\}_{i=1}^{4}$
are not the eigenstates of the Hamiltonian but the eigenstates of the LRI. In the interaction
picture, if the  environment is a vacuum reservoir, the instantaneous state must be a pure state,
i.e., $\ket{\psi_{3}(0)}$. When we select the initial state as this pure steady state,  the DMME in
the interaction picture (Eq.(\ref{eq:imeq})) ensures that the population on $\ket{\psi_{3}(0)}$ is
invariant. On the other  hand, since the unitary operator used to transform the picture that satisfies
Eq.(\ref{eq:ueqn}), the Hamiltonian used in the DMME in the Schrödinger picture is the Hamiltonian
given in Eq.(\ref{eq:Hs}). Therefore, we do not need to add a counterdiabatic Hamiltonian
to accelerate the adiabatic evolution \cite{Chen2011,Santos2021}.
Thus, a shortcut of the adiabatic evolution to connect initial and target eigenstate of the Hamiltonian
is established.

Now, let us consider the non-adiabatic control protocol from the initial separable state
$\ket{\phi_0}=\ket{00}$ to the maximal entanglement state
$\ket{\phi_T}=(\ket{00}-\ket{11})/\sqrt{2}$ by using the instantaneous steady state $\ket{\psi_{3}(t)}$.
For this purpose, we shall solve the set of the differential equations in Eq.(\ref{eq:geq}).
In order to determine the LRI Eq.(\ref{eq:Is}), we fix the boundary conditions by defining
\begin{eqnarray*}
g_{1}(0) & = &\frac{2\delta^{2}-1}{2\delta\sqrt{1-\delta^{2}}}\gamma,
\:g_{2}(0)=0,\:g_{6}(0)=\gamma,\\
g_{1}(T) & = & 0,\:g_{2}(T)=0,
\:g_{6}(T)=\frac{\mathrm{\gamma}}{2\delta\sqrt{1-\delta^{2}}},
\end{eqnarray*}
where $\delta$ and $\gamma$ are positive parameters with $\sin\eta_{2}(0)=\delta$ and
$\gamma\neq0$. Since the eigenvalues of the LRI are constants, it follows that $\lambda_{3}$,
$\delta$, and $\gamma$ are related by $\lambda_{3}=-\gamma
/\left(2\mathrm{\delta}\sqrt{1-\mathrm{\delta}^{2}}\right)$. Moreover, we observe that, as the
constant parameter $\delta$ tends to be zero, the steady state $\ket{\psi_{3}(t)}$ approaches to
$\ket{\phi_0}$ at $t=0$. Thus, we set following ansatz:
\begin{eqnarray}
g_{1}(t) & = & g_{1}(0)\sin^{2}\left(\omega_{e}t\right),\nonumber\\
g_{2}(t) & = & g_{2m}\sin(2\omega_{e}t),\nonumber\\
g_{6}(t) & = & \sqrt{\lambda_{3}^{2}-g_{1}^{2}(t)-g_{2}^{2}(t)},\label{eq:ansa}
\end{eqnarray}
with the control period $T=\pi/(2\omega_{e})$. $g_{2m}$ is a constant, which must be chosen
carefully to ensure a real $g_{6}(t)$.  Finally, employing the previous results, we can obtain the
functions $f(t)$ and $J(t)$ from Eq.(\ref{eq:geq}), which read
\begin{eqnarray}
&f(t)=\frac{\omega_{e}\left(2{g_{2m}^{2}}\cos\left(2\omega_{e}t\right)-{g_{1}^{2}(0)}
\cos^{2}\left(\omega_{e}t\right)\right)/4{g_{2m}}}{\sqrt{{g_{1}^{2}(0)}\left(1-\cos^{4}
\left(\omega_{e}t\right)\right)-{g}_{2m}^{2}\sin^{2}\left(2\omega_{e}t\right)+{g_{6}^{2}(0)}}},\nonumber\\
&J(t)=\frac{g_1(0)\omega_e}{2\pi g_{2m}}.\label{eq:jf}
\end{eqnarray}
Here we set $\hbar=1$ for simplifying our discussion.

\begin{figure}[htbp]
\centerline{\includegraphics[width=1.1\columnwidth]{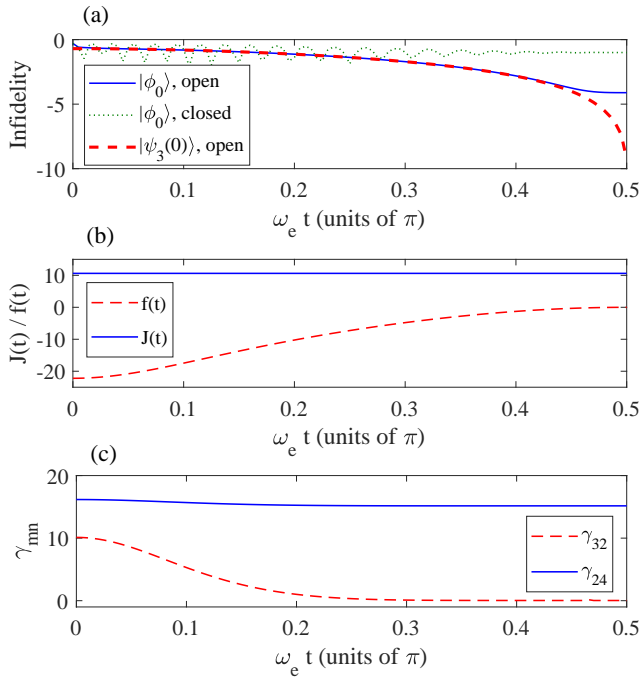}}
\caption{(a) The {infidelities} between $\rho_\text{s}(t)$ and $\ket{\phi_T}$ with a perfect initial state
$\ket{\psi_3(0)}$ (red dashed line) and a preset initial $\ket{\phi(0)}$ for both open
(blue solid line) and closed (green dotted line) cases, (b) Transverse field $f(t)$ and the scalar coupling
$J(t)$, (c) The decoherence rates $\gamma_{32}$ and $\gamma_{24}$ as functions of the
dimensionless time in the unity of $\pi$. The parameters are chosen as $\gamma=1$, $g_{2m}
=0.02$, $\delta=\sqrt{0.1}$, $\omega^c_{mn}=10\alpha_{mn}$, $s_{24}=0.01$ and $s_{32}=0.1$.
 We set $\omega_e=1$ as a unity of $f(t)$, $J(t)$, $\gamma_{32}$ and $\gamma_{24}$.}\label{fjg}
\end{figure}

For the inverse engineering method, the initial state is not $\ket{\phi_0}$,
but the initial eigenstate $\ket{\psi_{3}(0)}$. A small amplitude on $\ket{11}$ is required for
successful entanglement generation. Even so, because of the decoherence effect, employing
a not too large $\delta$, an initial state as $\ket{\phi_0}$ can still be transferred to the target
state with a satisfied fidelity. This will be shown in the following numerical results.

Next, we show numerical results for the Ohmic coupling spectral density function\cite{Benedetti2018}
\begin{eqnarray}
J(\alpha_{mn})=s_{mn}\alpha_{mn}\exp\left(-\frac{\alpha_{mn}}{\omega^c_{mn}}\right),\label{eq:sdf}
\end{eqnarray}
 where $s_{mn}$ denotes the dimensionless coupling strength and $\omega^c_{mn}$
is the cut-off frequency. In FIG.\ref{fjg} (a), we plot the infidelity $\log_{10}(1-\bra{\phi_T}
\rho_\text{s}(t)\ket{\phi_T})$ as a function of dimensionless time $\omega_e t$ for a perfectly
prepared initial state $\ket{\psi_3(0)}$ (the red  dashed line) and an imperfect initial state
$\ket{\phi_0}$ (the blue solid line). As a comparison, we also plot the infidelity for the closed
system case, where the evolution of the quantum state governed by Eq(\ref{eq:meq2}) with
$\gamma_{mn}=0$. By using the transverse field $f(t)$ and the scalar coupling  $J(t)$ as shown
in FIG.\ref{fjg} (b), the quantum state transfers into the target entanglement state with a
perfect fidelity (the red dashed line in FIG.\ref{fjg} (a)).  Obviously, the control protocol given by
the inverse engineering method based on the LRIs still works well in open dynamics case. But
the differences are also obvious: (i) The inverse engineering does not suit all of eigenstates of the
LRI, and only one of them can be used, which corresponds to the instantaneous steady state;
(ii) The inverse engineering can be {successful} only in the zero-temperature
reservoir, while it is a mixed state for the finite temperature case (see Eq.(\ref{eq:steadys})).

It should be noticed that, {because} the instantaneous frequency $\alpha_{mn}$ is time-dependent,
the target state may fail to generate, if the ansatz of $g_i (t)$ is not chosen well. For instance, if $g_{2m}$
exceeds a particular value (about 0.48 for the parameters chosen in FIG.\ref{fjg}), $\alpha_{32}$
will be negative {in some time ranges}, while $\alpha_{23}$ is becoming positive. {
Thus the transition direction between $\ket{\psi_2(t)}$ and $\ket{\psi_3(t)}$ is reversed.
In this time range, the  Lindblad operators involved in the DMME are $F_{23}(t)$ and
$F_{24}(t)$. The instantaneous steady  state of the DMME is no longer  $\ket{\psi_3(t)}$,
but $\ket{\psi_2(t)}$. Therefore, we will fail to prepare the target state $(\ket{00}-\ket{11})/\sqrt{2}$.
For the parameters used in FIG.\ref{fjg}, both $\alpha_{32}$ and $\alpha_{24}$ are  still
positive  which leads to positive decoherence rates as shown in FIG.\ref{fjg}(c).}

The ansatz given by Eq.(\ref{eq:ansa}) provides us with a constant scalar coupling and a decreasing transverse
field, which is very similar to the adiabatic engineering protocol.  {However, the trajectories from the initial state
to the target state are different. The adiabatic trajectory is defined by the instantaneous steady state of the
adiabatic master equation, i.e, $\ket{\psi_3(t)}$ with $\zeta_2=0$ (see Eq.(\ref{eq:eta2})). In contrast,
$\zeta_2(t)$ can be nonzero for the trajectory defined by the instantaneous steady state of the DMME.
This is essential for accelerating the adiabatic engineering process. According to Eqs.(\ref{eq:zeta})
and (\ref{eq:ansa}), $\zeta_2(t)$ will be zero if $g_{2m}=0$. Because $J(t),f(t)\propto g_{2m}^{-1}$
(see Eq.(\ref{eq:jf})),   the quantum state can evolve along the adiabatic trajectory by means of
either infinite $f$ and $J$,  or a infinitesimal $\omega_e$. Besides, a proper selection of
 $g_{2m}$ will  reduce the driving strength used in the quantum state engineering.}

On the other hand, it is shown by Eq.(\ref{eq:jf}) that the transverse field at $t=0$ satisfies
 \[f(0)=\frac{\omega_e(g_1^2(0)+2g_{2m}^2)}{4g_{2m}g_6(0)}.\]
Due to $\delta\rightarrow 0$, it yields $g_1(0)\rightarrow\infty$. Thus, if we prepare the initial
state on $\ket{00}$, the control protocol requires an infinite transverse field at $t=0$, which leads
to the inverse engineering scheme is unavailable. Therefore, we must admit a superposition state of
$\ket{00}$ and $\ket{11}$ as the initial state. However, it is difficult to prepare the initial state on
$\ket{\psi_3(0)}$ precisely. And the starting point of our control mission is not $\ket{\psi_3(0)}$, but
$\ket{\phi_0}=\ket{00}$. Due to the decoherence effect, we can always present a better fidelity than
the closed systems case. In FIG.\ref{fjg}(a), we plot the evolution of the fidelity for the initial
state $\ket{\phi_0}$, in which the blue solid line (the green dots line) associates with the mater
equation with (without) the dissipator. Since it is unitary evolution for the case without the
dissipator, the final fidelity will be 0.9 as shown by the green dotted line in FIG.\ref{fjg}(a).
In contrast to the closed systems case, the infidelity for the open system case decreases and
approaches to a perfect fidelity with the evolution. The decoherence draws the quantum system
into $\ket{\psi_3(t)}$ gradually. {Here, we would like to clarify that the inverse engineering
method and the  dissipation engineering method are totally different.  The inverse engineering
method is to transfer the quantum state from an initial eigenstate of the LRI to its final eigenstate,
which is usually the target state. But it is true only for the closed quantum systems.
For the open quantum system dynamics, the decoherence will draw the quantum state out the
eigenstate of the LRI.  In contrast, the dissipation engineering method is to generate
the target state by means of the decoherence effect. The target state is usually the steady
state of open quantum systems. Obviously, our proposal combines the advantages of
both two methods, and is robust to errors in  initial state preparing. Therefore,
the stronger decoherence rates are, the higher fidelity we can obtain within a finite
control period.}

At last, we would like to emphasise that the ansatz for $g_i$ for $i=1,2,6$
will be chosen optionally. For instance, the advantage of the ansatz given in
Eq.(\ref{eq:ansa}) is that scalar coupling $J$ is constant. Also we can choose
\[g_{1}(t)  =  g_{1}(0)\sin^{3}\left(\omega_{e}t\right).\]
In this way, the transverse field at $t=0$ reads
\[f(0)=\frac{{g_{2m}}\omega_e}{2\, {g_6(0)}},\]
which is independent on $g_1(0)$. Thus, even $\delta\rightarrow 0$, we can still have a
control protocol with a finite transverse field at $t=0$. Meanwhile, the scalar coupling
$J$ must be changed with time, which reads
\[J(t)=-\frac{3\, {g_1(0)}\omega_e}{4\, \pi\, \mathrm{g_{2m}}}\sin\left(\omega_e t\right).\]
If $g_{2m}/g_6(0)\ll 1$ is set, we will immediately obtain a  control protocol starting and
ending with zero transverse field.

\section{Conclusion} \label{sec:Conc}

In this paper, we explicitly solve the
problem of a double two-level system with a time-dependent transverse field and a scalar coupling
which interacts with a common heat reservoir in the finite temperature. By means of the
driven-Markovian master equation based on the Lewis-Riesenfeld invariants theory,
we show that both the decoherence rates
and the Lindblad operators are time-dependent, which implies a time-dependent steady state
will appear in the open system dynamics. Such a time-dependent steady state is an important
candidate in the quantum state engineering of open quantum systems. For instance, if the
reservoir is vacuum, the instantaneous steady state is not only a time-dependent pure state,
but also one of the LRI's eigenstates. Therefore, the quantum state can be transferred along
the trajectory given by this eigenstate with a perfect fidelity by means of the inverse engineering
method for the closed quantum systems, even if the initial state does not prepared precisely.
This implies a potential application in the non-adiabatic quantum control
by using the inverse engineering method based on the LRIs theory\cite{Kang2022,Whitty2022}.

As we see, the DMME depends on the driving protocol of the system. The generators of the Hamiltonian in the
DMME presented here belongs to a semi-simple subalgebras $so(4)\oplus u(1)$ of the Lie
algebras $su(4)$\cite{Nakahara2012}. For the other physical models and driving protocols for the double two-level
system, we need to analyse symmetry of driving protocol. Based on this symmetry and related
semi-simple subalgebras, the LRIs can be obtained explicitly. Fortunately, the LRIs for the
driven four-level system have been explored in Ref.\cite{Nakahara2012}. Therefore, following the procedure
presented in this paper, it is not difficult to derive the DMMEs for different driving protocols.

This work is supported by the National Natural Science Foundation of China (NSFC) under Grants
Nos. 12205037, 12075050 .

\appendix

\section{The Lamb shifts Hamiltonian} \label{A1}

We start with the DMME in the interaction picture (Eq.(\ref{eq:imeq})).
The Lamb shift Hamiltonian is given by Eq.(\ref{eq:lmh}). As shown in Sec. \ref{sec:REG},
there are two Lindblad operators involved in the DMME for the driven double two-level
system, i.e., $\tilde F_{32}=\ket{\psi_3(0)}\bra{\psi_2(0)}$ and $\tilde F_{24}=\ket{\psi_2(0)}
\bra{\psi_4(0)}$. Substituting the Lindblad operators into Eq.(\ref{eq:lmh}), we immediately
obtain the concrete Lamb shift Hamiltonian
\begin{eqnarray*}
 \tilde H_{\text{LS}}=&&\hbar S(\alpha_{32})\xi_{32}^{2}\ket{\psi_2(0)}\bra{\psi_2(0)}
\nonumber\\&&+\hbar S(\alpha_{24})\xi_{24}^{2}\ket{\psi_4(0)}\bra{\psi_4(0)}.
\end{eqnarray*}
Since $\alpha_{mn}$ and $\xi_{mn}$ are time-varying, the Lamb shift Hamiltonian in the
interaction picture is a time-dependent operator. When the reservoir is at zero temperature,
i.e., $N_{mn}=0$, $S(\alpha_{mn})$ can be given analytically. Considering a  Ohmic
coupling spectral density function as shown in Eq.(\ref{eq:sdf}), we have
\begin{eqnarray*}
S(\alpha_{mn})=\frac{s_{mn}}{\hbar^2}\left[\omega_c-\alpha_{mn}\exp\left(-\frac{\alpha_{mn}}{\omega_c}\right)
\text{Ei}\left(\frac{\alpha_{mn}}{\omega_c}\right)\right],
\end{eqnarray*}
where $\text{Ei}(x)=\int_{-\infty}^x\ e^{-x'}/x' dx'$ is the one-argument exponential integral
function. Thus  the Lamb shift Hamiltonian in the Schr\"{o}dinger picture can be obtained by
means of $ H_{\text{LS}}=U_\text{s} \tilde H_{\text{LS}}U_\text{s} ^\dagger$.

In the following, we verify that $\ket{\psi_3(t)}$ is the instantaneous steady state, or the
dark state, of the DMME with the Lamb shifts. We still focus on the DMME in the interaction
picture at first
\begin{eqnarray}
\partial_{t}{\tilde\rho}_{\text{s}}&\equiv&\mathcal{\tilde L}\tilde{\rho}_{\text{s}}(t)=
-\frac{i}{\hbar}\left[\tilde H_{\text{LS}}(t),\tilde\rho_{\text{s}}\right]\nonumber\\
 & +&\gamma_{32}\left(\tilde F_{32}{\tilde\rho}_{\text{s}}\tilde F_{32}^{\dagger}-\frac{1}{2}
 \left\{\tilde F_{32}^{\dagger}\tilde F_{32},{\tilde\rho}_{\text{s}}\right\} \right)\nonumber\\
 & + &\gamma_{24}\left(\tilde F_{24}{\tilde\rho}_{\text{s}} \tilde F_{24}^{\dagger}-\frac{1}{2}
 \left\{\tilde F_{24}^{\dagger}\tilde F_{24},{\tilde\rho}_{\text{s}}\right\} \right).\label{eq:a1}
\end{eqnarray}
If the steady state in the interaction picture is $\ket{\psi_3(0)}$ , the instantaneous steady of the
DMME in the Schr\"{o}dinger picture must be $\ket{\psi_3(t)}$  owing to $\rho_\text{ss}(t)=
U_\text{s} \tilde \rho_\text{ss}U_\text{s} ^\dagger$. Here we use the criteria of the pure steady state
given in Ref.\cite{Kraus2008}. It is not difficult to see that $\ket{\psi_3(0)}$ satisfies the following
conditions, i.e., $\tilde H_{\text{LS}}(t)\ket{\psi_3(0)}=0$, $\tilde F_{32}\ket{\psi_3(0)}=0$ and
$\tilde F_{24}\ket{\psi_3(0)}=0$. Therefore, we confirm that $\ket{\psi_3(0)}$ is the steady state of
the DMME Eq.(\ref{eq:a1}). On the other hand, if we discard the Lamb shifts Hamiltonian, the steady
state is not changed, so is the instantaneous steady state in the Schr\"{o}dinger picture
($\ket{\psi_3(t)}$).

\begin{figure}[htbp]
\centerline{\includegraphics[width=1.1\columnwidth]{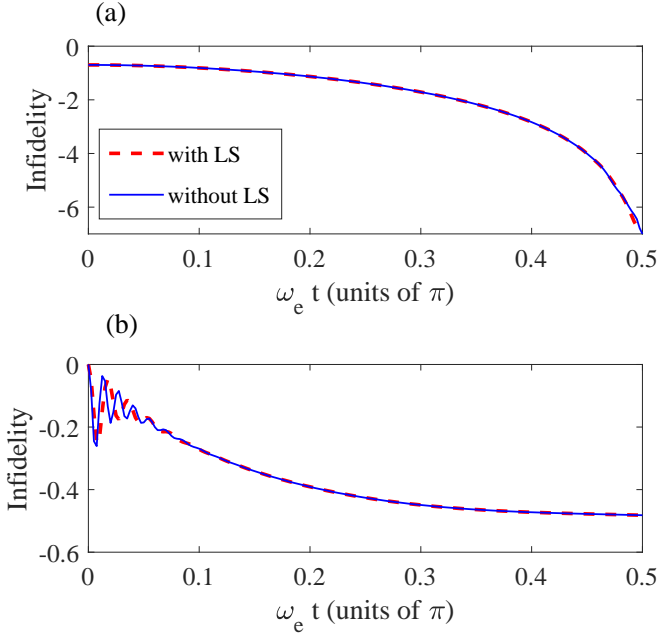}}
\caption{(a) The infidelities between $\rho_\text{s}(t)$ and $\ket{\phi_T}$ with the Lamb shifts
 (red dashed line) and without the Lamb shifts (blue solid line) for the initial state $\ket{\psi_3(0)}$,
  (b) The infidelities between $\rho_\text{s}(t)$ and $\ket{\phi_T}$ with the Lamb shifts
 (red dashed line) and without the Lamb shifts (blue solid line) for the initial state $\ket{\psi_4(0)}$.
 The parameters are chosen as $\gamma=1$, $g_{2m}
=0.02$, $\delta=\sqrt{0.1}$, $\omega^c_{mn}=10\alpha_{mn}$, $s_{24}=0.01$ and $s_{32}=0.1$.
 We set $\omega_e=1$ as a unity of $f(t)$, $J(t)$, $\gamma_{32}$ and $\gamma_{24}$.}\label{ls}
\end{figure}

In order to show this visibly, we recheck the infidelity between the quantum states,
governed by the DMME with and without the Lamb shifts, and the target state $\ket{\Phi_T}$, which
is shown in FIG.\ref{ls}. When the initial state is chosen as the initial steady state $\ket{\psi_3(0)}$,
the dynamical evolutions governed by the DMME with and without the Lamb shifts are similar,
which is verified by FIG.\ref{ls} (a).
Therefore, we can neglect the Lamb shift terms in the DMME if we are interested in generating
a quantum state by the instantaneous steady state. Otherwise, the Lamb shifts may affect the dynamical
evolution when we choose the other initial states.  In FIG.\ref{ls} (b), we also plot the infidelity
with the initial state $\ket{\psi_4(0)}$. As we see, the dynamical evolutions are  evidently different
at the beginning. With the evolution, they decay into the same steady state $\ket{\psi_3(t)}$.
Therefore, we can still trust the numerical results in Sec. \ref{sec:REG}, where there is a tiny
population out of the initial steady state.

\end{document}